  \documentclass[%
   reprint,
  superscriptaddress,
  showpacs,showkeys,preprintnumbers,
   amsmath,amssymb,
   aps,
  ]{revtex4-1}

\usepackage{graphicx}
\usepackage{dcolumn}
\usepackage{bm}
\usepackage{here}
\usepackage[dvips]{color}
\usepackage{multirow}
\usepackage{ulem}
\usepackage[columnwise]{lineno}

\makeatletter
  \def\erf{\mathop{\operator@font erf}\nolimits}
  \def\erfc{\mathop{\operator@font erfc}\nolimits}
  \def\Erf{\mathop{\operator@font Erf}\nolimits}
  \def\Shi{\mathop{\operator@font Shi}\nolimits}
  \def\Chi{\mathop{\operator@font Chi}\nolimits}
  \def\Ei{\mathop{\operator@font Ei}\nolimits}
  \def\cosec{\mathop{\operator@font cosec}\nolimits}
  \def\sech{\mathop{\operator@font sech}\nolimits}
  \def\cosech{\mathop{\operator@font cosech}\nolimits}
  \newcommand\hypgeo[2]{{}_{#1}{\operator@font F}_{#2}}
  \def\Re{\mathop{\operator@font Re}\nolimits}
  \def\Im{\mathop{\operator@font Im}\nolimits}
\makeatother

\begin{document}


\title{Transverse asymmetry of individual $\bm{\gamma}$-rays in the $^{139}$La($\bm{\vec{n}}$, $\bm{\gamma}$)$^{140}$La reaction}

\def\affNagoya{Nagoya University, Furocho, Chikusa, Nagoya 464-8602, Japan}
\def\affIbaraki{Ibaraki University, 2-1-1 Bunkyo, mito, Ibaraki 310-8512, Japan}
\def\affIndiana{Indiana University, Bloomington, Indiana 47401, USA}
\def\affKyushu{Kyushu University, 744 Motooka, Nishi, Fukuoka 819-0395, Japan}
\def\affTohoku{Tohoku University, 2-1-1 Katahira, Aoba, Sendai, Miyagi 980-8577, Japan}
\def\affJAEA{Japan Atomic Energy Agency, 2-1 Shirane, Tokai 319-1195, Japan}
\def\affTokyoTech{Tokyo Institute of Technology, Meguro, Tokyo 152-8551, Japan}
\def\affRCNP{Osaka University, 10-1 Mihogaoka, Ibaraki, Osaka 567-0047, Japan}
\def\affKEK{KEK, 1-1 Oho, Tsukuba, Ibaraki 305-0801, Japan}

\author{M.~Okuizumi}
\affiliation{\affNagoya}

\author{C.~J.~Auton}
\affiliation{\affIndiana}
\affiliation{\affNagoya}

\author{S.~Endo}
\affiliation{\affNagoya}
\affiliation{\affJAEA}

\author{H.~Fujioka}
\affiliation{\affTokyoTech}

\author{K.~Hirota}
\affiliation{\affKEK}

\author{T.~Ino}
\affiliation{\affKEK}

\author{K.~Ishizaki}
\affiliation{\affNagoya}

\author{A.~Kimura}
\affiliation{\affJAEA}

\author{M.~Kitaguchi}
\affiliation{\affNagoya}

\author{J.~Koga}
\affiliation{\affKyushu}

\author{S.~Makise}
\affiliation{\affKyushu}

\author{Y.~Niinomi}
\affiliation{\affNagoya}

\author{T.~Oku}
\affiliation{\affJAEA}
\affiliation{\affIbaraki}

\author{T.~Okudaira}
\affiliation{\affNagoya}
\affiliation{\affJAEA}

\author{K.~Sakai}
\affiliation{\affJAEA}

\author{T.~Shima}
\affiliation{\affRCNP}

\author{H.~M.~Shimizu}
\affiliation{\affNagoya}

\author{H.~Tada}
\affiliation{\affNagoya}

\author{S.~Takada}
\affiliation{\affTohoku}
\affiliation{\affJAEA}

\author{S.~Takahashi}
\affiliation{\affIbaraki}

\author{Y.~Tani}
\affiliation{\affTokyoTech}

\author{T.~Yamamoto}
\affiliation{\affNagoya}

\author{H.~Yoshikawa}
\affiliation{\affRCNP}

\author{T.~Yoshioka}
\affiliation{\affKyushu}


\date{\today}


\begin{abstract}
The enhancement of the parity-violating asymmetry in the vicinity of $p$-wave compound nuclear resonances was observed for a variety of medium-heavy nuclei. 
The enhanced parity-violating asymmetry can be understood using the $s$-$p$ mixing model.
The $s$-$p$ mixing model predicts several neutron energy-dependent angular correlations between the neutron momentum $\vec k_n$, neutron spin $\vec\sigma_n$, $\gamma$-ray momentum $\vec k_\gamma$, and $\gamma$-ray polarization $\vec\sigma_\gamma$ in the $(n,\gamma)$ reaction. 
In this paper, the improved value of the transverse asymmetry of $\gamma$-ray emissions, corresponding to a correlation term $\vec{\sigma}_n\cdot(\vec k_n\times\vec k_\gamma)$ in the $^{139}\mathrm{La}(\vec n,\gamma)^{140}\mathrm{La}$ reaction, and the transverse asymmetries in the transitions to several low excited states of $^{140}\mathrm{La}$ are reported.
\end{abstract}

\maketitle


\section{Introduction}
A large parity-violating asymmetry of the cross section has been observed in a $p$-wave resonance of the $^{139}\mathrm{La} + n$ compound state~\cite{Alfimenkov}. 
The magnitude of the parity-violating asymmetry amounts to $10^6$ times larger than that of nucleon-nucleon scattering, which is dominated by the interference between parity-unfavored partial waves via the contribution of the weak interaction in the compound nuclear process~\cite{pot74,yua86,ade85,Shimizu1993}.
The enhanced parity-violating asymmetry is explained as the result of the interference between the $p$-wave resonance and neighboring $s$-wave resonances ($s$-$p$ mixing model)~\cite{sus82}.
The $s$-$p$ mixing model predicts angular correlations between neutron momentum, neutron spin, $\gamma$-ray momentum, and $\gamma$-ray spin depending on neutron energy in $(n,\gamma)$ reactions~\cite{fla85}. 
Investigating these spin angular correlations is essential for understanding the parity violation enhancement mechanism.

Angular correlations in the $(n,\gamma)$ reaction of several nuclei have been observed for the integral $\gamma$-spectrum~\cite{Vesna2015}.
Recently, the neutron energy dependence of the angular distribution was measured for individual $\gamma$-rays emitted from $p$-wave resonances in $^{139}$La~\cite{Okudaira18, Okudaira21}, $^{117}$Sn~\cite{Koga22}, and $^{132}$Xe~\cite{Okudaira23} using an intense pulsed neutron beam at the Material and Life Science Experimental Facility (MLF) of the Japan Proton Accelerator Research Complex (J-PARC).
The transverse asymmetry was also measured in  $^{139}$La~\cite{Yamamoto20, Yamamoto20e} and $^{117}$Sn~\cite{Endo22}.
The transverse asymmetry corresponds to the correlation term $a_2\vec\sigma_n\cdot(\vec k_n\times \vec k_\gamma)$ in Ref.~\cite{fla85}, where $\vec\sigma_n$, $\vec k_n$, and $\vec k_\gamma$ are unit vectors for the neutron spin, neutron momentum, and $\gamma$-ray momentum direction, respectively.

We have accumulated more statistics with the setup described in Ref.~\cite{Yamamoto20, Yamamoto20e}.
The transverse asymmetry was measured with improved statistics for $\gamma$-ray transition to the ground state and also low-excited states.
\section{Experiment and Analysis}
\subsection{Experiment}

The measurements were carried out using the Accurate Neutron-Nucleus Reaction Measurement Instrument (ANNRI) instrument on beamline-04 of MLF at J-PARC~\cite{ANNRI}. 
The neutron energy was determined using the neutron time-of-flight (TOF) method.
The neutron beam was polarized using a $^3$He spin-filter, which makes use of the large spin dependent cross-section of $^3$He nuclei.

The $^3$He spin was flipped approximately every four hours using the adiabatic fast-passage NMR method~\cite{Ino12}. 
The $\gamma$-ray energy was measured with germanium detectors. 
Li-Glass detectors were installed downstream of the target, which are used for neutron transmission measurements.

The ANNRI instrument uses two types of Ge detectors: cluster-type and coaxial-type detectors, each positioned to surround the target in both the vertical and horizontal directions~\cite{Takada18}.
For this analysis, only the up- and down-cluster detectors were utilized.
Further details on the experimental setup and measurements are described in Ref.~\cite{Yamamoto20}.

\begin{figure}[htb]
	\begin{center}
        \includegraphics[width=0.99\linewidth]{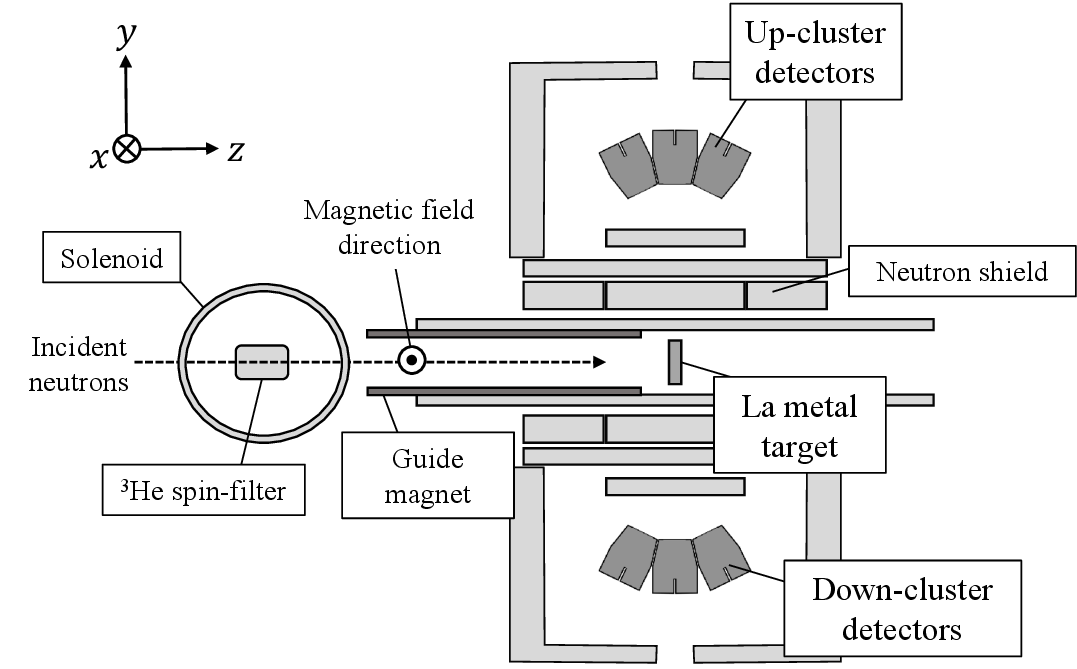}
	\caption[]{
    The sectional view around the target. There are seven Ge detectors above and below the target.
}
	\label{fig:setup}
	\end{center}
\end{figure}

A metal lanthanum plate of $40~\mathrm{mm}\times40~\mathrm{mm}\times3~\mathrm{mm}$ with a purity of 99.9\% was used as the nuclear target.
The setup around the target is shown in Fig.~\ref{fig:setup}.
The total measurement time was 222 hours.

\subsection{Analysis}

\begin{figure}[htb]
	\centering
	\includegraphics[width=0.9\linewidth]{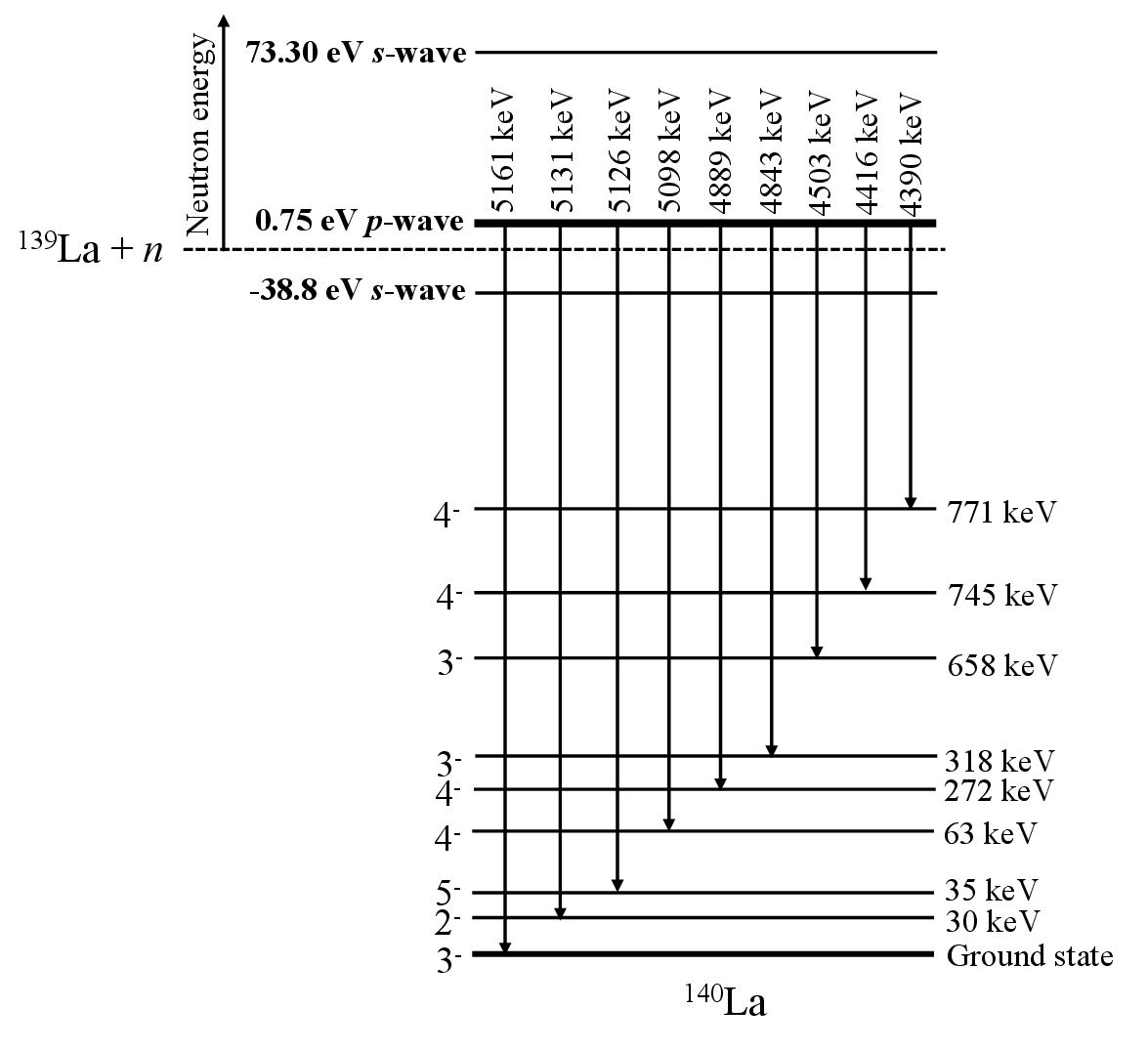}
	\caption[]{
    Transitions from $^{139}\mathrm{La} + n$ resonance state to the low excited states and the ground state of $^{140}\mathrm{La}$. 
    The resonance energy values are taken from Ref.~\cite{endo2023measurements}
  }
	\label{scheme}
\end{figure}

\begin{figure}[htb]
	\centering
	\includegraphics[width=1.1\linewidth]{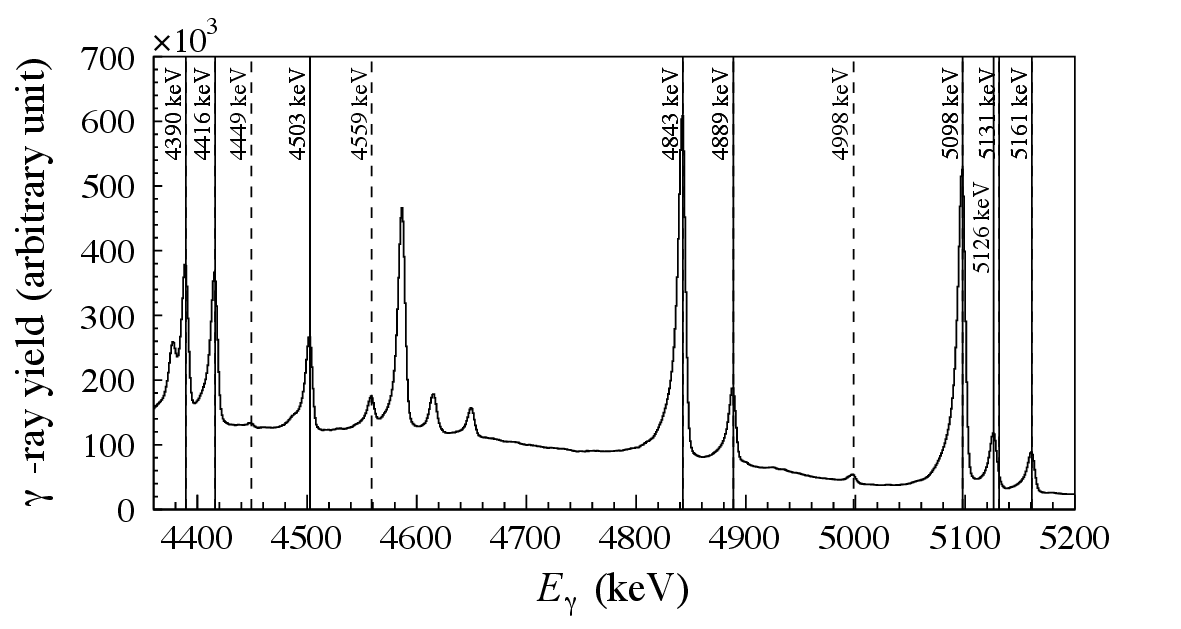}
	\caption[]{
  The peaks indicated by solid and dashed lines are full absorption peaks of $^{139}$La.
  We focused on the eight peaks indicated by solid lines.
  Peaks indicated by the dashed line were not analyzed due to an insufficient number of events.
  Three peaks around 4600 keV and on the left side of 4390 keV peak are the single escape peaks from peaks of 5161 keV, 5131 keV, 5126 keV, 5098 keV, and 4889 keV peaks.
  }
	\label{gamma}
\end{figure}

This analysis focuses on the $\gamma$-rays emitted in each transition to final states with excitation energies of $E_\gamma = 0,~30~\text{or}~35,~63,~272,~318,~658,~745$, and $771~\mathrm{keV}$.
The transition scheme from $^{139}\mathrm{La}+n$ to $^{140}\mathrm{La}$ is depicted in Fig.~\ref{scheme}. 
Figure~\ref{gamma} shows the $\gamma$-ray spectrum corresponding to the energy range of transitions to low excited states in $^{140}$La. 
The peaks at 5126 keV and 5131 keV cannot be separated and were treated as a single peak.
Events within the full-width at the quater-maximum region of the eight photo peaks indicated by the solid lines in Fig.~\ref{gamma} were used for the analysis.

\begin{figure*}[htb]
	\centering
	\includegraphics[width=0.9\linewidth]{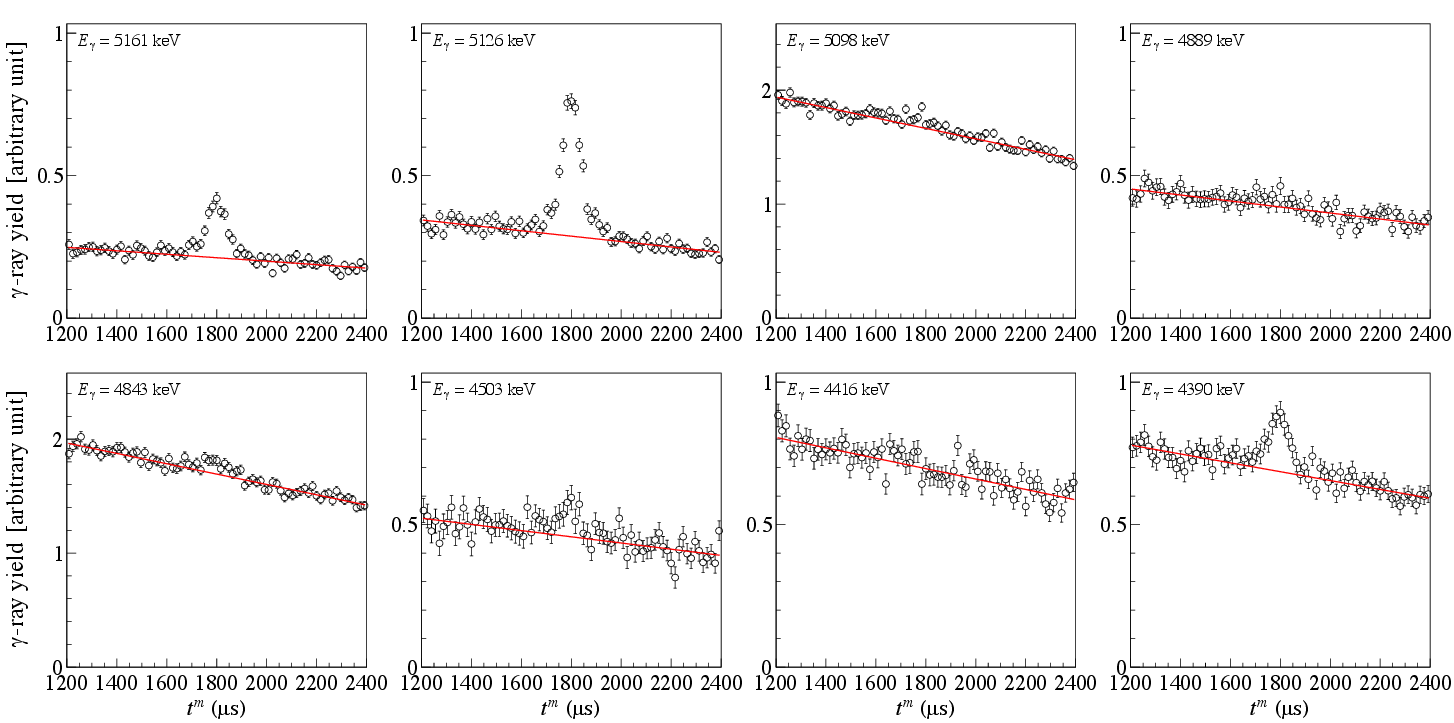}
	\caption[]{
$\gamma$-ray yield in the vicinity of the $p$-wave resonance and the result of background fitting. The fitted spectra were averaged for both spin directions.
The energy values of photopeaks are shown in the top left of each histogram.
  }
	\label{tofBG}
\end{figure*}

The pileup rate of the $\gamma$-ray signal was less than 5\%, which was corrected for.
The detection efficiency of each germanium detector was determined from measurements of the $^{14}\mathrm{N}(n,\gamma)^{15}\mathrm{N}$ reaction, which emits $\gamma$ rays isotropically.
The Compton scattering background was estimated using a third-order polynomial fit on both sides of each peak.
The $s$-wave resonance components were subtracted by fitting with a linear function to obtain the asymmetry of the $p$-wave resonance. 
Figure~\ref{tofBG} shows the TOF spectra after the corrections with a fit of the $s$-wave component.

\begin{figure*}[htbp]
	\centering
        \includegraphics[width=0.9\linewidth]{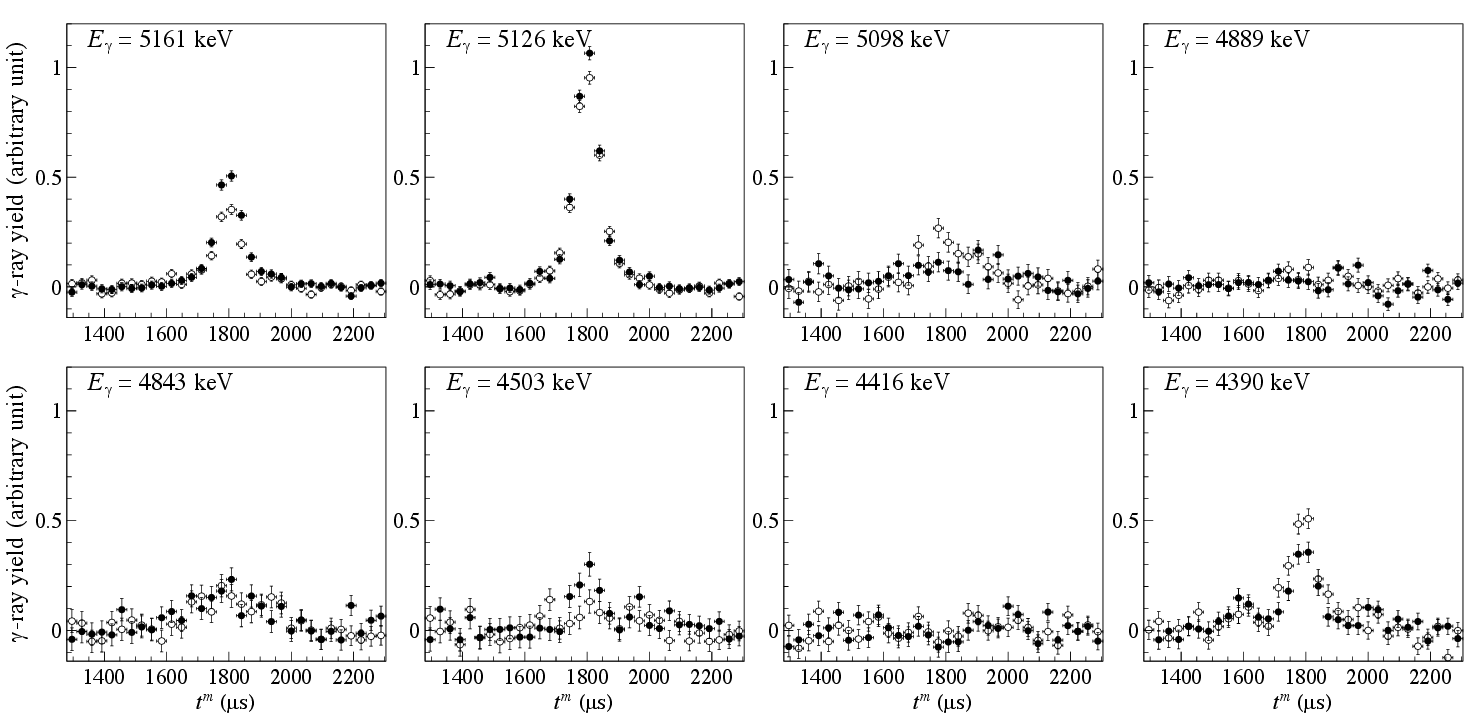}
	\caption[]{
    $\gamma$-ray yield in the vicinity of the $p$-wave resonance for each spin direction.
  White points indicate $n^\mathrm{up}_-+n^\mathrm{down}_+$ which correspond to positive sign of $\vec{\sigma}_n\cdot(\vec k_n\times\vec k_{\gamma})$ and black points indicate $n^\mathrm{up}_++n^\mathrm{down}_-$ which correspond to negative sign of $\vec{\sigma}_n\cdot(\vec k_n\times\vec k_{\gamma})$.
  }
	\label{tof_updown}
\end{figure*}

In a right-hand coordinate system, we define the $z$-axis  corresponding to the neutron beam direction. 
The neutron spin polarization is along the positive or negative $x$-axis direction. 
We label the spin direction along the positive and negative $x$-axis direction as "+" and "-" respectively.
We define $n^\mathrm{up}_\pm$ and $n^\mathrm{down}_\pm$ as the corrected $\gamma$-ray yields obtained with up- and down- cluster detectors.
The summed TOF spectra of the $p$-wave region corresponding to each sign of $\vec{\sigma}_n\cdot(\vec k_n\times\vec k_{\gamma})$ are shown in Fig.~\ref{tof_updown}.

\subsection{Transverse Asymmetry}
The asymmetry is defined as
\begin{eqnarray}
  \epsilon^\mathrm{up,down}_\gamma=\frac{N^\mathrm{up,down}_+-N^\mathrm{up,down}_-}{N^\mathrm{up,down}_++N^\mathrm{up,down}_-}.
\end{eqnarray}
where, $N^\mathrm{up}_\pm$ and $N^\mathrm{down}_\pm$ are the integral of the corrected TOF spectra with a range $E_r - 2\Gamma_p\leq E_n \leq E_r + 2\Gamma_p$ obtained from up- and down-cluster detectors respectively, where $\Gamma_p=\Gamma^\gamma+\Gamma^n$. 
The resonance parameters of $^{139}$La+$n$ are listed in Table~\ref{tab:res}.
\begin{table}[htbp]
\begin{center}
	\begin{tabular}{ccccc}
	\hline
	\hline
	$E_r\,[{\rm eV}]$ & $J$ & $l$ & $\Gamma^{\gamma}\,[{\rm meV}]$ & $g\Gamma^n\,[{\rm meV}]$ 	\\
	\hline
	$0.750\pm 0.001$ & $4$ & $1$ & $41.6\pm 0.9$ & $(3.67 \pm0.5 )\times 10^{-5}$ \\
	\hline
	\hline
	\end{tabular}
	\caption{
    Resonance parameters of the $^{139}\mathrm{La} + n$ $p$-wave resonance state. 
    $E_r,~J,~l,~\Gamma^\gamma,~g,$ and $\Gamma^n$ are the resonance energy, total angular momentum of the resonance state, orbital angular momentum of the incident neutron, $\gamma$ width, $g$ factor, and neutron width, respectively.
    The values are taken from Ref.~\cite{endo2023measurements}.\\
	}
	\label{tab:res}
\end{center}	
\end{table}

The neutron polarization is determined from the $^3$He polarization, which was obtained from measurements of the neutron transmission rate. 
The corrected transverse asymmetry with neutron polarization is expressed as:

\begin{eqnarray}
  {A'}^\mathrm{up,down}_\mathrm{LR}=\frac{2\epsilon^\mathrm{up,down}_{\gamma}}{(P^+_n+P^-_n)-\epsilon^\mathrm{up,down}_\gamma(P^+_n-P^-_n)},
\end{eqnarray}
where $P_n^+$ and $P_n^-$ are the average neutron polarization during measurements for each spin direction.

The effect of scattering in the target was corrected as:
\begin{eqnarray}
  A_\mathrm{LR}^\mathrm{up,down}={A'}^\mathrm{up,down}_\mathrm{LR}\left(1+\frac{N_{s\geqslant1}}{N_{s=0}}\right),
\end{eqnarray}
where $N_{s\geq1}$ is the number of events in which neutrons scattered in the target and were subsequently absorbed by the nucleus. 
$N_{s=0}$ is the number of events that emit $\gamma$ rays without scattering.
The ratio of these quantities was estimated by Monte Carlo simulation in Ref.~\cite{Yamamoto20}.

We determined the transverse asymmetry for each up and down-cluster for each measurement, and their weighted average was used as the final result $A_\mathrm{LR}$.
Note that the sign of $A^\mathrm{up}_\mathrm{LR}$ is inverted when averaging since $A^\mathrm{up}_\mathrm{LR}$ corresponds to $-\vec{\sigma}_n\cdot(\vec k_n\times\vec k_{\gamma})$ and $A^\mathrm{down}_\mathrm{LR}$ corresponds to $\vec{\sigma}_n\cdot(\vec k_n\times\vec k_{\gamma})$.
The results of $A_\mathrm{LR}$ are shown in Table~\ref{tab:result}.
The transverse asymmetry for the transition to the ground state was obtained with over $5\sigma$ significance.
Non-zero transverse asymmetries were found in the transitions to the excited states of 63 keV and 772 keV with a confidence level of over 99.7\%.

\begin{table}[htbp]
\begin{center}
	\begin{tabular}{lccc}	
	\hline
	\hline
	$E_{\gamma}$ [keV] &
	$E_{\rm ex}$ [keV]& 
	$F$ & 
  $A_\mathrm{LR}$ 
	\\
  \hline
5161 & 0 & 3 &               $ -0.86 \pm 0.10 $\\
5131, 5126 & 30, 35 & 2,5 &  $ -0.08 \pm 0.05 $\\
5098 & 63 & 4 &              $ 1.43 \pm 0.39 $\\
4889 & 272 & 4 &             $ 0.21 \pm 0.83 $\\
4843 & 318 & 3 &             $ -0.01 \pm 0.37 $\\
4503 & 658 & 3 &             $ -1.04 \pm 0.53 $\\
4416 & 745 & 4 &             $ -0.17 \pm 4.21 $\\
4390 & 771 & 4 or 5 or 6 &         $ 0.90 \pm 0.19 $\\
  \hline
	\hline
	\end{tabular}
	\caption{
    The final state spin $F$ and the result of $A_\mathrm{LR}$ for each photopeak.
	}
	\label{tab:result}
\end{center}	
\end{table}


\section{Conclusion}
We measured the transverse asymmetries for several transitions from the $p$-wave resonance in $^{139}\mathrm{La}+n$ to several low-excited states of $^{140}$La.
Transverse asymmetries were observed for transitions to three final states with a confidence level of over 99.7\%.
There are indications of angular correlations for other transitions.
The consistency of the $a_1$ and $a_2$ terms~\cite{fla85} was discussed in Ref.~\cite{Nakabe23} using measurement results for $^{139}$La~\cite{Okudaira18,Yamamoto20,Yamamoto20e,OkudairaImB}. The same analysis will be performed using the improved values in this paper, and the consistency will be discussed for the transitions to the excited states.

\begin{acknowledgments}
The authors would like to thank the staff of ANNRI for the maintenance of the germanium detectors, and MLF and J-PARC for operating the accelerators and the neutron-production target. The neutron experiments at the Materials and Life Science Experimental Facility of J-PARC were performed under the user program (Proposals No. 2018B0148 and No. 2019A0185). This work was supported by the Neutron Science Division of KEK as an S-type research project with program number 2018S12. This work was partially supported by MEXT KAKENHI Grant No. JP19GS0210 and JSPS KAKENHI Grant Nos. JP17H02889, 20K14495, and 23K13122. 
\end{acknowledgments}

\bibliography{ngamma}

\end{document}